# Observation of exciton redshift-blueshift crossover in monolayer $WS_2$

E. J. Sie,[1] A. Steinhoff,[2] C. Gies,[2] C. H. Lui,[3] Q. Ma,[1] M. Rösner,[2,4] G. Schönhoff,[2,4] F. Jahnke,[2] T. O. Wehling,[2,4] Y.-H. Lee,[5] J. Kong,[6] P. Jarillo-Herrero,[1] and N. Gedik*[1]

[1]Department of Physics, Massachusetts Institute of Technology, Cambridge, Massachusetts 02139, United States
[2]Institut für Theoretische Physik, Universität Bremen, P.O. Box 330 440, 28334 Bremen, Germany
[3]Department of Physics and Astronomy, University of California, Riverside, California 92521, United States
[4]Bremen Center for Computational Materials Science, Universität Bremen, 28334 Bremen, Germany
[5]Materials Science and Engineering, National Tsing-Hua University, Hsinchu 30013, Taiwan
[6]Department of Electrical Engineering and Computer Science, Massachusetts Institute of Technology, Cambridge, Massachusetts 02139, United States

*Corresponding Author: gedik@mit.edu

**Abstract:**

We report a rare atom-like interaction between excitons in monolayer $WS_2$, measured using ultrafast absorption spectroscopy. At increasing excitation density, the exciton resonance energy exhibits a pronounced redshift followed by an anomalous blueshift. Using both material-realistic computation and phenomenological modeling, we attribute this observation to plasma effects and an attraction-repulsion crossover of the exciton-exciton interaction that mimics the Lennard-Jones potential between atoms. Our experiment demonstrates a strong analogy between excitons and atoms with respect to inter-particle interaction, which holds promise to pursue the predicted liquid and crystalline phases of excitons in two-dimensional materials.

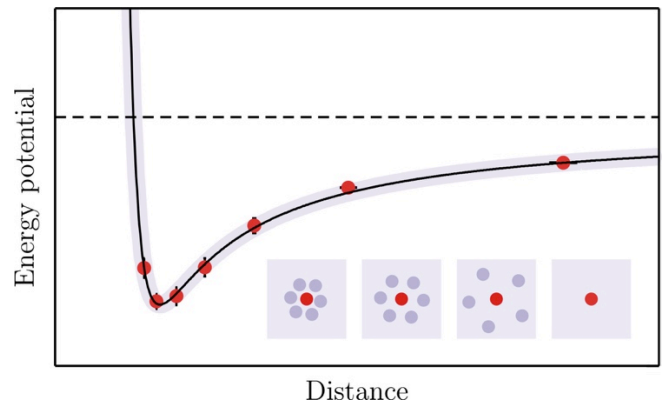

Table of Contents Graphic

**Main Text:**

Excitons in semiconductors are often perceived as the solid-state analogs to hydrogen atoms. This analogy helps us to understand the basic features of excitons, notably their internal energy states. However, this analogy breaks down as we consider the inter-particle interactions because of some fundamental differences between atoms and excitons. Atoms are stable particles with large ionization energy (~10 eV). They exhibit long-range van der Waals attraction and short-range Pauli repulsion, which form the so-called Lennard-Jones potential as a function of interatomic separation [1, 2]. In contrast, excitons are transient quasiparticles with much smaller binding energy and extremely short lifetime. They can dissociate into an electron-hole plasma, whose relative concentration is governed by the law of mass action [3, 4]. Hence, interactions in semiconductors are somewhat different from those in real gases, because the effects from plasma [5] and excitons [6] must be considered. The relative importance of exciton and plasma effects depends on the regime of excitation density. In conventional III-V and II-VI semiconductors, such complex many-body effects preclude the demonstration of atom-like interactions between excitons, particularly in the regime of high excitation density where excitons become unstable.

Recent advances in two-dimensional (2D) semiconductors, particularly monolayer transition-metal dichalcogenides (TMDs), offer a unique platform to investigate excitonic interactions. These materials possess strong Coulomb interactions due to quantum confinement and reduced dielectric screening [7], leading to the formation of excitons with exceptionally large binding energies ~300 meV [8-11]. The enhanced stability of excitons in these materials provides good opportunities to reexamine the role of plasma effects and excitonic interactions over a broad range of excitation density. Although the exciton physics in photoexcited TMDs has been much studied [12-17], a complete picture of excitonic interactions in these materials is still lacking.

In this Letter, we investigate systematically the many-particle interactions in monolayer $WS_2$. We combine ultrafast absorption spectroscopy, microscopic many-body theory and an analytic approach that maps the measured exciton-exciton interactions onto an effective atomic model. In particular, we measure the absorption spectrum of the A exciton under femtosecond optical excitation. As we increase the excitation density, we observe a pronounced redshift (73 meV) of the exciton resonance energy at low density, followed by an unusual blueshift (10 meV) at high density. We attribute the two different energy shifts to two distinct interaction regimes. At high density, the exciton blueshift is well described by assuming a repulsive exciton-exciton interaction similar to the short-range Lennard-Jones interaction between atoms [1, 2]. At low density, the Lennard-Jones potential further contains a long-range contribution due to an attractive exciton-exciton interaction. However, in contrast to the atomic case, the exciton redshift in this regime is found to follow a strongly modified exponent, indicating that not all carriers are bound excitons but a fraction exists as electron-hole plasma. Insight from microscopic theory reveals that the redshift observed at low excitation density mainly arises from plasma-induced bandgap renormalization and screening of the exciton binding energy. We note that the observed energy shift is much larger than those reported in conventional semiconductors such as GaAs quantum wells (~0.1 meV) [18-22] as a consequence of the much enhanced many-body interactions in the 2D material. Our model is further supported by the temperature dependent exciton energy shift observed in the time-resolved absorption measurements.

We investigate monolayer $WS_2$ samples grown on sapphire substrates by chemical vapor deposition [23-25]. In our pump-probe experiment, we generate carriers by femtosecond pump pulses with photon energy of 3.16 eV, well above the quasiparticle bandgap (~2.3 eV) [11]. We then monitor the A exciton resonance near 2.0 eV by recording the reflection spectrum of broadband probe pulses with controlled time delay at room temperature (Fig 1a). For a monolayer sample on a transparent substrate, the absorption spectrum can be extracted from the reflection spectrum using the thin-film approximation [14, 26, 27]. We estimate the excitation density $n$ from the measured incident pump fluence and absorbance of the sample at the excitation wavelength [27].

Figure 1b shows the absorption spectra of the A exciton at increasing pump fluence up to 18 μJ/cm$^2$ ($n = 5.3 \times 10^{12}$ cm$^{-2}$). The spectra were taken at a pump-probe delay of 2 ps, a time after which the excitons are expected to have reached thermal equilibrium with one another and with the lattice but not yet recombined [28, 29]. All the spectra can be fitted well with a Lorentzian function plus a second-order polynomial function (smooth lines over the data curves), which represent the exciton peak and the background, respectively [26]. We note that the B exciton is well separated from the A exciton by 400 meV in monolayer $WS_2$ [27], and therefore does not affect our analysis. From the fitting, we extract the exciton peak energy ($E_A$), linewidth ($\Gamma$), peak intensity ($I$) and spectral weight ($S$, i.e. the integrated area), and plot their changes in Fig 1c-e. While the spectral weight remains unchanged at all excitation densities, the other quantities vary significantly with the density.

These quantities exhibit two distinct behaviors at low and high excitation densities. At low density, the peak energy redshifts gradually for ~70 meV as the density increases, while the linewidth and peak intensity remain almost constant. The energy shift cannot be explained using the pump-induced lattice heating because the estimated lattice temperature increase is ~20 K, which corresponds to merely ~4 meV of temperature dependent gap narrowing [30]. In contrast, with further increase of density $n > 2.0 \times 10^{12}$ cm$^{-2}$, the rate of exciton redshift diminishes and eventually turns into a blueshift of ~10 meV from the lowest energy point. Notably, the redshift-blueshift crossover is accompanied by a large spectral broadening, with the linewidth increasing to more than twice of the initial width (Fig 1d). Correspondingly, the peak intensity drops to one half of the initial intensity to maintain the total spectral weight of the A exciton (Fig 1e). These two distinct energy shifts correspond to two different interaction regimes as we will discuss in the following.

We first discuss the redshift at low density. According to prior studies in monolayer TMDs [5], the exciton redshift can be ascribed to a combination of bandgap renormalization and plasma screening of the exciton binding energy due to the excited unbound carriers (Fig 2a). We have obtained the quasiparticle band structure and Coulomb matrix elements of monolayer WS$_2$ by first-principle G$_0$W$_0$ calculations. Screening from the substrate is additionally incorporated in the Coulomb matrix elements [31]. The exciton shift due to the excited carriers is calculated with microscopic semiconductor Bloch equations in a screened-exchange Coulomb-hole approximation (SXCH) – see Supporting Information [26]. With increasing excitation density, both the quasiparticle band gap ($E_g$) and the exciton binding energy ($E_b$) are found to decrease (Fig 2b). Since $E_g$ decreases faster than $E_b$, the resulting exciton resonance energy ($E_A = E_g - E_b$) shifts to lower energies (Fig 2c).

Our calculations reproduce the measured redshift at low density (Fig 1c, purple curve). The overall agreement is remarkable, given that we do not use any fitting parameter in our theory. We note that the SXCH calculation predicts a Mott transition at $n = 2 \times 10^{12}$ cm$^{-2}$. Approximately at this density, the experimental data reveals a crossover into an anomalous blueshift (Fig 1c), which we attribute to exciton-exciton interaction that is facilitated by the increasing fraction of carriers bound into excitons and which marks the limit in carrier density to which a plasma picture applies.

Although the observed redshift at low density is dominated by the plasma contribution, an additional contribution also results from exciton-exciton attraction as in the case of atomic van der Waals forces. Mutual attraction can reduce the energy cost to create an extra exciton ($E_A$) by a magnitude as much as the negative inter-exciton potential energy. In light of this picture, we interpret the blueshift at high density > 8 uJ/cm$^2$ ($n > 2.0 \times 10^{12}$ cm$^{-2}$), where a large fraction of carriers form bound excitons, as arising from an exciton-exciton *repulsion*. Indeed, the blueshift is not captured by our numerical approach, and therefore suggests a new contribution other than the plasma effects. The simultaneous broadening of the absorption peak indicates that this new interaction strongly perturbs the excitons and shortens their lifetime (Fig 1d). In this scenario, high-density excitons tend to repel each other due to the Pauli exclusion of overlapping electron orbitals, giving rise to positive inter-exciton potential energy. As a consequence, the energy cost to create an extra exciton increases, leading to a blueshift of the resonance energy.

The above interpretation has inspired us to quantify the excitonic contributions to the energy shift in the entire density range through a simple phenomenological model with two power laws in analogy to the well-known Lennard-Jones potential between atoms (for which $k = 6$):

$$\Delta E = \varepsilon \left[ \left(\frac{r_0}{r_s}\right)^8 - \left(\frac{r_0}{r_s}\right)^k \right] \qquad (1)$$

Here $r_s$ is the radius of disk occupied by an exciton ($n\pi r_s^2 = 1$). $\varepsilon$, $r_0$ and $k$ are the fitting parameters, which can be interpreted in a similar way as in the usual '12-6' power-law potential between atoms. The first term describes the exciton blueshift caused by the short-range Pauli repulsion. We use the $r_s^{-8}$ functional form for better fitting of the Pauli repulsion in this system instead of the usual $r_s^{-12}$ typically chosen in atomic system for convenience due to the relative computing efficiency ($r_s^{-12}$ is the square of $r_s^{-6}$). The second term models the exciton redshift caused by the long-range van der Waals attraction of excitons behaving as fluctuating dipoles in the presence of plasma in this material. In general, the functional form of this attraction potential can differ from the usual London dispersion force $r_s^{-6}$, hence we parameterize it as $r_s^{-k}$. By fitting the $\Delta E$–$r_s$ data through the least squares method with $\varepsilon = 128 \pm 10$ meV, $r_0 = 2.6 \pm 0.1$ nm and $k = 1.4 \pm 0.2$, our simple model matches precisely the density dependence of exciton energy shift (Fig 3a). Note that we have considered the effect of exciton-exciton annihilation in the determination of the estimated carrier densities (Supporting Information). This effect has been shown to play a role in highly excited monolayer TMDs [28, 29, 32, 33]. We note that the exponent $k$ differs (down to one fifth) from the exponent in the Lennard-Jones potential due to the presence of plasma effects in semiconductors, which are not captured by the atomic model. The obtained $r_0$ = 2.6 nm represents the exciton Bohr radius in monolayer WS$_2$. This value agrees well with our calculated exciton radius (2.3 nm) [26] and the estimated radius (1–3 nm) in other studies [11, 34, 35].

Excitons have been perceived as the solid-state counterpart of atoms, but the analogy is usually drawn only for their similar internal structure and molecular structures. The latter is apparent from the formation of trions [36-38] and biexcitons [14, 15, 39-41] with binding energies of 20-60 meV in monolayer TMDs that are analogous to the hydrogen anion and hydrogen molecules. Here, the good agreement between the modified Lennard-Jones model and our experiments reveals further that they also share similar mutual interaction behavior at long and short distances. This finding is remarkable because high-density excitons are usually unstable against the electron-hole plasma formation and other annihilation processes. These competing processes can easily destroy the exciton resonance features and hinder the observation of inter-exciton repulsion. Monolayer WS$_2$ is, however, an exceptional material, which hosts tightly-bound excitons with radius approaching the atomic limit. The robustness of these excitons helps maintain their resonance features even at very high density. We can therefore observe an effective attraction-repulsion crossover of excitonic interactions, a phenomenon that was predicted early [42] but remained unobserved experimentally until now.

Although the observed shift mimics the Lennard-Jones potential, there are three features distinct from atoms that deserve more careful attention. First, in addition to excitons, plasma can be present simultaneously with a relative density governed by the law of mass action. The interparticle separation $r_s$ is derived from their combined densities. Second, plasma contribution to the shift at low density dominates the exciton contribution. The observed $\sim 1/r_s$ dependence, instead of $1/r_s^6$, indicates a negligible contribution from excitonic van der Waals attraction. This is not surprising because excitons in monolayer TMDs are tightly bound. Third, a possible formation of biexcitons at short distance is not explicitly captured in this model. This is a similar situation faced by Lennard-Jones potential because it does not explicitly represent chemical bonding between atoms, but it can explain why a cluster of atoms can form at the minimum potential. Although biexciton formation may occur at the minimum potential (Fig 3a), the strong plasma screening precludes such occurrence. This is evidenced from the much reduced exciton binding energy at such a high excitation density (Fig 2b). This means biexciton binding energy should also be reduced to a value much smaller than the reported values, or completely screened, and thus unlikely to form.

We can further test our model (Eq. 1) through its temperature-dependent behavior. Two effects arise when the exciton temperature is high. First, such highly energetic excitons will be in constant motion and dynamically average out their short-range and long-range interactions amongst each other. Secondly, the excitons will have higher probability to reach their internal excited states. These effects will reduce the effective potential energy and increase their Bohr radius. As a consequence, the energy potential well between the excitons will become *shallower* and the inter-exciton distance at the potential minimum will become *wider* (inset of Fig 4c). In other words, when the excitons cool from very high to low temperature, we predict a significant redshift of their energy. As we discuss below, our model captures the complex cooling dynamics and offers a simple interpretation based on such exciton picture, although rigorous contribution from plasma effects could be included for a more accurate model.

This prediction can be conveniently explored in the cooling dynamics of excitons after pump excitation at high density regime, where exciton-exciton interaction dominates and the exciton picture is particularly appropriate. As we pump monolayer $WS_2$ using 3.16 eV photons ($h\nu > E_A$), we create free electron-hole pairs that immediately form excitons within the excitation pulse duration [43, 44]. This is due to the strong Coulomb attraction in monolayer $WS_2$ that leads to a very rapid exciton formation. The large excess energy will bring the excitons to a high temperature ($T_e > 1000$ K). Subsequently, we can follow the time evolution of the exciton resonance as they cool down. Figures 4a-b show a time series of the exciton absorption spectra up to 1 ns ($F = 11$ uJ/cm$^2$), from which we extract their peak parameters (Fig 4c-d). By examining these parameters, the exciton dynamics can be described roughly in three stages.

In Stage I (red region), the hot excitons are formed with rapidly decreasing inter-exciton distance, accompanied by a dramatic energy redshift and spectral broadening. In Stage II (blue), the hot excitons cool to the lattice temperature via phonon emission. In Stage III (yellow), the excitons recombine gradually with increasing inter-exciton distance leading to a blueshift. In the framework of our model, such exciton dynamics can be adequately described by the trajectory (I→II→III) between the hot and cold potential energy curves, as illustrated in the inset of Fig 4c. We exclude possible contribution from lattice cooling in Stage III because the estimated temperature-dependent energy shift is merely ~4 meV (20 K), far too small than the observed shift ~70 meV (Supporting Information). It is particularly noteworthy to examine the dynamics in Stage II, where the excitons cool down from $T_e > 1000$ K to ~300 K but the density should remain unchanged. As depicted by arrow II in the inset of Fig 4c, the exciton energy decreases when the high-temperature potential curve (red) evolves into the low-temperature curve (blue). During this process, the exciton energy is found to redshift for about 20 meV, accompanied by a decrease of linewidth. With this interpretation, we can assign to Stage II an exciton cooling time of 2 ps, comparable with the cooling time measured in graphene [45].

In summary, we have observed a transition of inter-excitonic interaction at increasing exciton density in monolayer $WS_2$, which manifests as a redshift-blueshift crossover of the exciton resonance energy. At low density, the exciton redshift arises from plasma screening effects and the long-range exciton-exciton attraction. At high density, the exciton blueshift is attributed to the short-range exciton-exciton repulsion. We describe this density dependence of the excitonic interactions by a phenomenological model, in analogy to the Lennard-Jones interaction between atoms, combined with a material-realistic computation of plasma effects. Interpreting the time dependence of energy shifts shortly after the carrier excitation in terms of our model, we extract an exciton cooling time of about 2 ps. Similar results are also observed in monolayer $MoS_2$, implying that this behavior is ubiquitous in monolayer TMD semiconductors [14]. The close analogy between the excitons and atoms, as shown in our experiment, suggests that the liquid and crystal phases of excitons [46-48] can be realized in 2D materials.

**Acknowledgments:**

We thank Timm Rohwer for discussion and carefully reading our manuscript. N.G. and E.J.S acknowledge support from the U.S. Department of Energy, BES DMSE (experimental setup and data acquisition) and from the Gordon and Betty Moore Foundation's EPiQS Initiative through Grant GBMF4540 (data analysis and manuscript writing). Work in the P.J.H. group work was partly supported by the Center for Excitonics, an Energy Frontier Research Center funded by the U.S. Department of Energy (DOE), Office of Science, Office of Basic Energy Sciences under Award Number DE-SC0001088 (measurement) and partly through AFOSR grant FA9550-16-1-0382 (data analysis), as well as the Gordon and Betty Moore Foundation's EPiQS Initiative through Grant GBMF4541 to P.J.H. J.K. acknowledges support from STC Center for Integrated Quantum Materials, NSF Grant DMR-1231319 (material growth). Y-H.L. acknowledges support AOARD grant (co-funded with ONRG) FA2386-16-1-4009, Ministry of Science and Technology (MoST 105-2112-M-007-032-MY3; MoST 105-2119-M-007-027), and Academia Sinica Research Program on Nanoscience and Nanotechnology, Taiwan (material growth). The Bremen groups acknowledge financial support from the Deutsche Forschungsgemeinschaft via the research training group "*Quantum Mechanical Materials Modeling*". M.R., G.S. and T.W. thank the European Graphene Flagship for financial support.

**Notes:**

The authors declare no competing financial interests.

**Supporting Information:**

Experimental methods; Kramers-Kronig analysis; Maxwell's equations for monolayer materials; Fitting analysis; Microscopic many-body computation; Exciton-exciton annihilation effect; Heat capacity and estimated temperature.

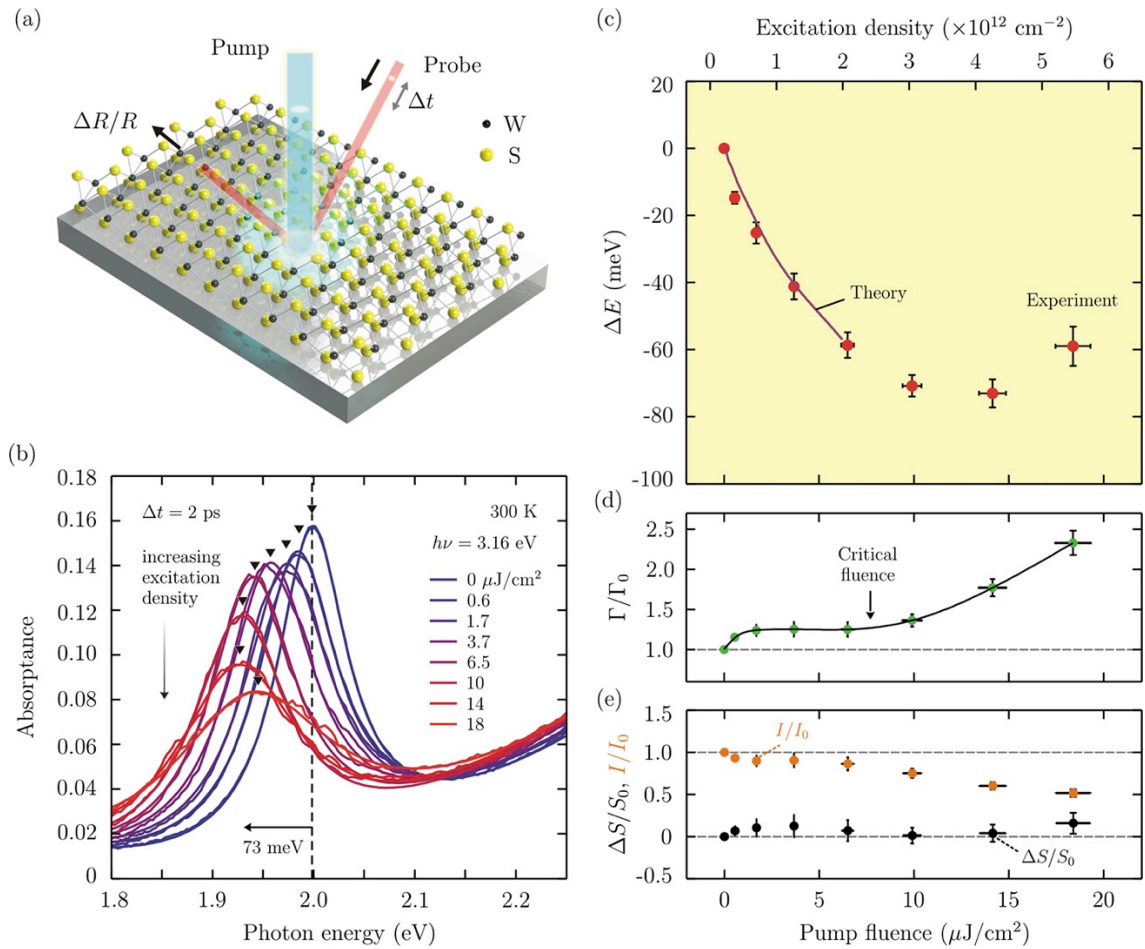

**Figure 1.** (a) Schematic of transient absorption spectroscopy setup. (b) Absorption peak of A exciton in monolayer $WS_2$ at increasing excitation densities. The fitting lines (smooth lines) are superimposed with the data curves. (c) Exciton energy shift ($\Delta E$) as a function of pump fluence obtained from experiment (red) and calculation (purple). The calculated value terminates at $n = 2\times10^{12}$ cm$^{-2}$ due to a Mott transition that is predicted prematurely by the SXCH theory. (d) Linewidth ($\Gamma/\Gamma_0$), (e) peak height ($I/I_0$) and spectral weight change ($\Delta S/S_0$).

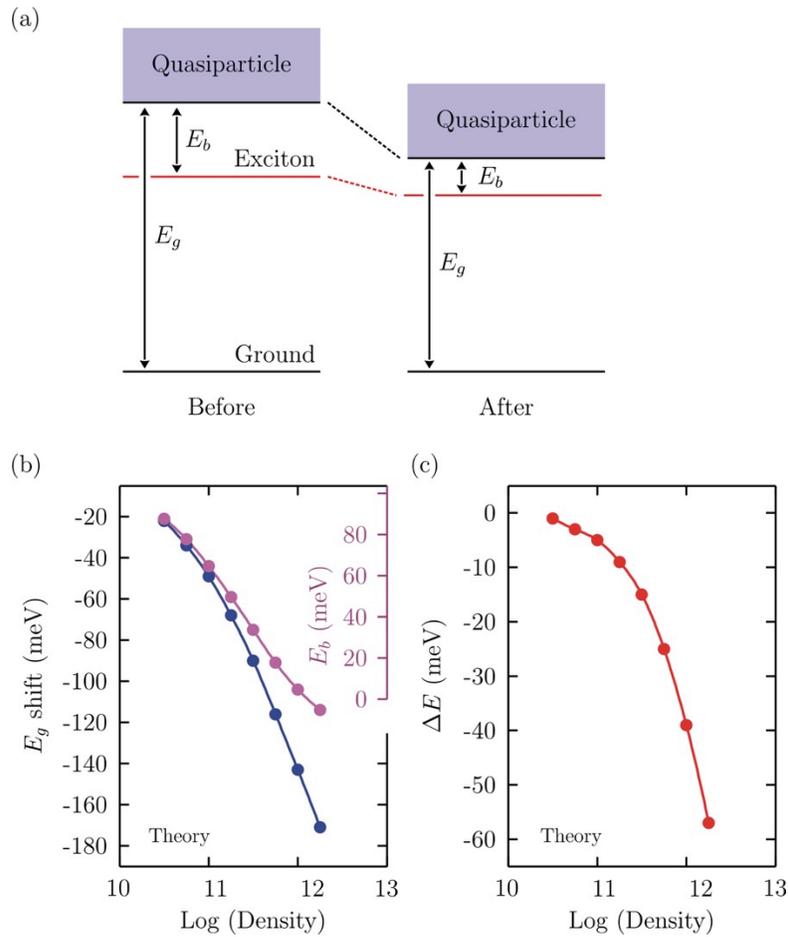

**Figure 2.** (a) Schematic band diagrams before and after the femtosecond optical pumping. Both the band gap ($E_g$) and the exciton binding energy ($E_b$) are reduced after photoexcitation, giving rise to a net downshift of exciton resonance energy. (b) Calculated bandgap narrowing ($\Delta E_g$) and exciton binding energy ($E_b$) as a function of exciton density in a semi-logarithmic scale. (c) The resultant shift of exciton resonance energy ($\Delta E$) at increasing exciton density, as also shown in Figure 1c.

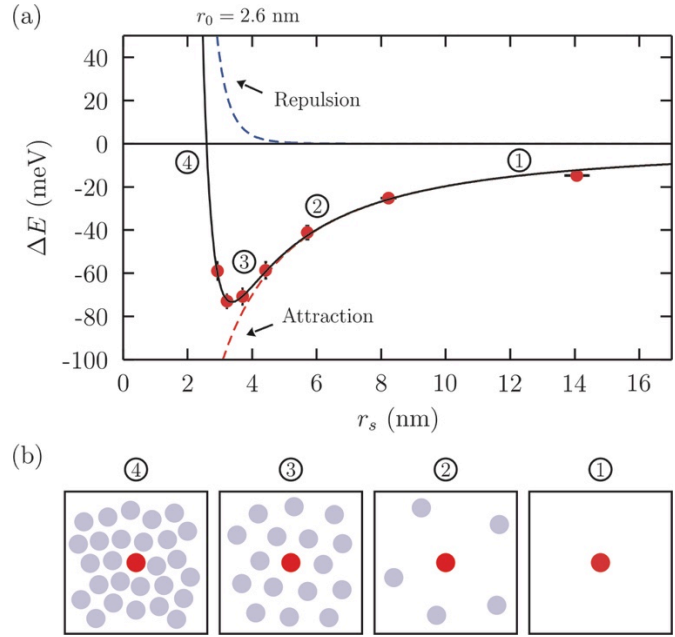

**Figure 3.** (a) Exciton energy shift ($\Delta E$) as a function of average radius ($r_s$) occupied by an exciton in the exciton gas. The red dots are experimental data from Figure 1c. The solid black line is the best fit of our phenomenological model (Eq. 1). The dashed lines are the repulsion and attraction components of the inter-exciton potential. $r_0 = 2.6$ nm is the extracted exciton radius. (b) Schematic configuration of a probe exciton (red) among the pump-generated excitons (purple) at different interaction regimes (1-4) as denoted in panel (a).

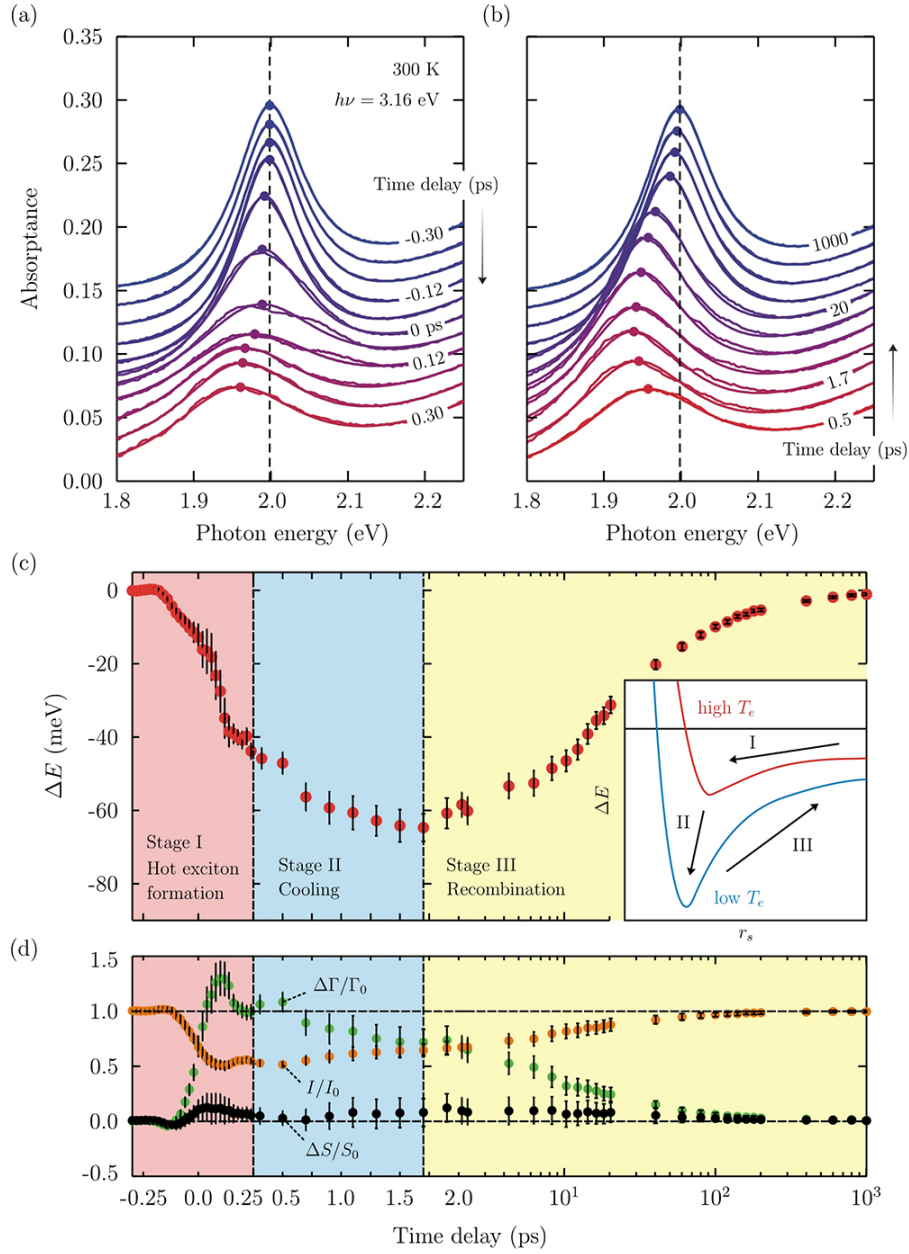

**Figure 4.** (a, b) Exciton absorption peak at increasing time delays from -0.30 to 0.30 ps (a) and from 0.5 to 1000 ps (b), where only the subset of the data is shown for clarity. The fitting lines (smooth lines) are superimposed with the data curves. (c) Exciton energy shift ($\Delta E$), as well as (d) peak height ($I/I_0$), linewidth broadening ($\Delta \Gamma/\Gamma_0$), and spectral weight change ($\Delta S/S_0$). Inset in (c) shows the stages of relaxation dynamics through the $\Delta E$ vs $r_s$ curve at high and low temperatures.

# Supporting Information for

# Observation of exciton redshift-blueshift crossover in monolayer WS$_2$

E. J. Sie,[1] A. Steinhoff,[2] C. Gies,[2] C. H. Lui,[3] Q. Ma,[1] M. Rösner,[2,4] G. Schönhoff,[2,4] F. Jahnke,[2] T. O. Wehling,[2,4] Y.-H. Lee,[5] J. Kong,[6] P. Jarillo-Herrero,[1] and N. Gedik*[1]

[1]Department of Physics, Massachusetts Institute of Technology, Cambridge, Massachusetts 02139, United States
[2]Institut für Theoretische Physik, Universität Bremen, P.O. Box 330 440, 28334 Bremen, Germany
[3]Department of Physics and Astronomy, University of California, Riverside, California 92521, United States
[4]Bremen Center for Computation Materials Science, Universität Bremen, 28334 Bremen, Germany
[5]Materials Science and Engineering, National Tsing-Hua University, Hsinchu 30013, Taiwan
[6]Department of Electrical Engineering and Computer Science, Massachusetts Institute of Technology, Cambridge, Massachusetts 02139, United States

*Corresponding Author: gedik@mit.edu

1. **Experimental methods**
2. **Kramer-Kronig analysis**
3. **Maxwell's equations for monolayer materials**
4. **Fitting analysis**
5. **Microscopic many-body computation**
6. **Exciton-exciton annihilation effect**
7. **Heat capacity and estimated temperature**

## 1. Experimental methods

The sample consists of high-quality monolayers of $WS_2$ that were CVD-grown on sapphire substrates [1, 2], and all measurements in this study were conducted at ambient condition (300 K, 1 atm). In our experiments, we used a Ti:sapphire amplifier producing laser pulses with duration of 50 fs and at 30 kHz repetition rate. Each pulse was split into two arms. For the pump arm, the pulses were sent to a second-harmonic nonlinear crystal, while for the probe arm the pulses were sent through a delay stage and a white-light continuum generator ($h\nu = 1.78\text{-}2.48$ eV, chirp-corrected). The two beams were focused at the sample, and the probe beam was reflected to a monochromator and a photodiode for lock-in detection [3, 4]. By scanning the grating and the delay stage, we were able to measure $\Delta R/R$ (and hence $\alpha$, [3]) as a function of energy and time delay.

## 2. Kramers-Kronig analysis

Throughout our analysis, we used the proper definitions of reflectance $R$, transmittance $T$ and absorptance $\alpha$, which respectively means the fractions of incident electromagnetic power that is reflected, transmitted and absorbed at the monolayer interface (between vacuum and substrate). This is in contrast to reflectivity and transmittivity, which are technically only valid for semi-infinite system. In the main text, we also used more familiar names such as absorbance and absorption to mean, quantitatively, the absorptance $\alpha$ of monolayer $WS_2$.

Pump-probe experiments detect small changes in probe reflectance (or transmittance) that is induced by pump excitation. This gives the differential reflectance $\Delta R/R$ as a function of energy and time delay, from which we can obtain the transient reflectance, $R(t) = R_0(1 + \Delta R(t)/R_0)$, where $R_0$ is the reflectance of the system in equilibrium. In fact, the absorptance $\alpha$ (or the induced absorptance $\Delta\alpha$) is what we really want (as shown in the main text) because it provides the explicit information about the optical transition matrix element of the system. The absorptance and the reflectance are related through the complex dielectric function $\tilde{\varepsilon}$. This relation can be derived using Maxwell equations (see section S3). We obtain $\tilde{\varepsilon}(\omega, t)$ by fitting $R(\omega, t)$ using a Kramers-Kronig (KK) constrained variational analysis [5]. Finally, we construct $\alpha(\omega, t)$ by repeating this procedure at different time delays. The details of the above procedure are described as follows.

First, we want to find the relation between the complex dielectric function and the optical properties such as reflectance, transmittance and absorptance by using Maxwell's equations. It is important to include the substrate influence on electromagnetic radiation especially for atomically-thin materials. Here, the current density in a monolayer $WS_2$ sample is described by a delta function, $j_x = \tilde{\sigma}(\omega)\delta(z)E_x$ where $\tilde{\sigma}$ is the complex conductivity and $E_x$ is the $x$-component of the probe electric field (along the sample's surface). By substituting this into Maxwell's equations and using the appropriate boundary conditions between the monolayer and the substrate, we can obtain the reflectance as

$$R(\omega) = \frac{(1 - n_s - \frac{\omega d}{c}\epsilon_2)^2 + (\frac{\omega d}{c}(\epsilon_1 - 1))^2}{(1 + n_s + \frac{\omega d}{c}\epsilon_2)^2 + (\frac{\omega d}{c}(\epsilon_1 - 1))^2} \tag{1}$$

and the transmittance as

$$T(\omega) = \frac{4n_s}{(1 + n_s + \frac{\omega d}{c}\epsilon_2)^2 + (\frac{\omega d}{c}(\epsilon_1 - 1))^2} \tag{2}$$

where $n_s$ is the substrate's refractive index (1.7675 for sapphire at photon energy of 2.07 eV), $d$ is the effective thickness of the monolayer (0.67 nm), $\epsilon_1$ and $\epsilon_2$ are the real and imaginary parts of the dielectric function, respectively. Here, the 2D dielectric function is expressed as

$$\tilde{\epsilon}(\omega) = 1 + \frac{4\pi i \tilde{\sigma}/d}{\omega} \tag{3}$$

Meanwhile, the absorptance can be expressed as

$$\alpha(\omega) = \frac{4\frac{\omega d}{c}\epsilon_2}{(1 + n_s + \frac{\omega d}{c}\epsilon_2)^2 + (\frac{\omega d}{c}(\epsilon_1 - 1))^2} \tag{4}$$

These expressions are exact, and they are valid for any monolayer materials on a dielectric substrate. We find that the presence of the substrate significantly influences the optical properties of the monolayer WS$_2$ above it. As compared to an isolated monolayer WS$_2$, the reflectance is enhanced, while both the transmittance and the absorptance are reduced. In graphene, the above expressions can be further simplified because the real part of its dielectric function is featureless in the visible spectrum ($\epsilon_1 \sim 1$, negligible $\sigma_2$). This is, however, not the case for monolayer WS$_2$, and we must include both the real and imaginary parts of the dielectric function to obtain accurate results. In situation where none of the equilibrium absorptance, reflectance or transmittance spectrum is available, the pump-induced absorptance $\Delta\alpha$ can still be estimated pretty well from the measured $\Delta R/R$ or $\Delta T/T$ through the following expression

$$\frac{\Delta R}{R} = \left[\left(\frac{n_s + 1}{n_s - 1}\right) + \frac{n_s}{(n_s - 1)^2}\frac{(\gamma_1^2 + \gamma_2^2)}{\gamma_2}\right]\Delta\alpha \tag{5}$$

$$\frac{\Delta T}{T} = -\left[\left(\frac{n_s + 1}{2}\right) + \frac{(\gamma_1^2 + \gamma_2^2)}{4\gamma_2}\right]\Delta\alpha \tag{6}$$

where $\Delta R$ and $\Delta T$ are the pump-induced changes of the probe reflectance $R$ and transmittance $T$, $n_s$ is the refractive index of the substrate, $\gamma_1 = \omega d(\epsilon_1 - 1)/c$, and $\gamma_2 = \omega d \epsilon_2 / c$. In situation where $\gamma_1^2$ and $\gamma_2^2$ are small, such as graphene and few other TMDs, only the first term in the bracket needs be considered.

In our analysis, we used the equilibrium absorptance $\alpha$ of monolayer WS$_2$ measured using differential reflectance microscopy (see main text). The absorptance spectrum contains peaks from the A exciton at 2.0 eV. The equilibrium reflectance $R_0$ can then be constructed from $\alpha$ by finding the appropriate complex dielectric function $\tilde{\epsilon}$ as expressed in equations (3), (4) and (1). To do this, we implemented a Kramers-Kronig (KK) constrained variational analysis [5] to extract $\tilde{\epsilon}$ from the measured $\alpha$ in thin-film approximation. Here, the total dielectric function is constructed by many Drude-Lorentz oscillators, which are anchored at equidistant energy spacing, in the following form

$$\tilde{\epsilon}(\omega) = \epsilon_\infty + \sum_{k=1}^{N} \frac{\omega_{p,k}^2}{\omega_{0,k}^2 - \omega^2 - i\omega\gamma_k} \tag{7}$$

In our calculations, we used $N = 40$ oscillators with a fixed linewidth of $\gamma_k = 50$ meV spanning the energy range of $1.77 \; eV \leq \omega_{0,k} \leq 2.40 \; eV$, and we found that these parameters can fit $\alpha$ spectrum very well. We can then construct $R_0$ spectrum by using $\tilde{\epsilon}$ obtained from the above analysis.

The transient absorptance spectra $\alpha(t)$ can be obtained by performing similar (KK) analysis. This time we inferred the absorptance from the reflectance at different time delays: $R(t) = R_0(1 + \Delta R(t)/R_0)$, where the differential reflectance $\Delta R(t)/R_0$ is measured directly from the experiments.

3. Maxwell's equations for monolayer materials

In this section, we provide a full derivation from Maxwell's equations in order to obtain the exact solutions of reflectance $R(\omega)$, transmittance $T(\omega)$ and absorptance spectra $\alpha(\omega)$ for any monolayer materials on a substrate. Readers who are interested in this section should also refer to the original articles by L. A. Falkovsky [6] and T. Stauber [7] in the study of graphene. Here, we express the solutions in terms of the complex dielectric function $\tilde{\epsilon}(\omega)$ or conductivity $\tilde{\sigma}(\omega)$ for Kramers-Kronig analysis as shown in section S2.

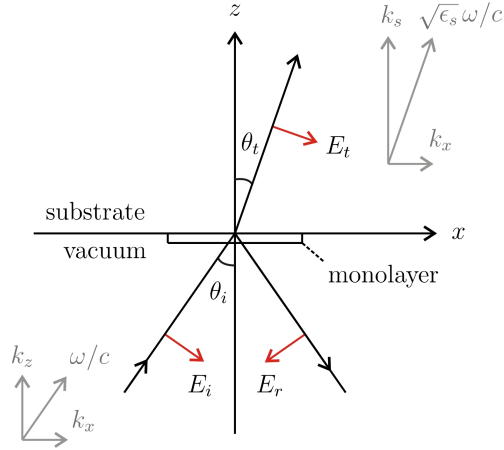

**Figure S1.** Schematic of the plane of incidence.

The electromagnetic wave equation at frequency $\omega$, inside a medium with a dielectric constant $\epsilon$ and a current density **j**, can be expressed as

$$\nabla(\nabla \cdot \mathbf{E}) - \nabla^2 \mathbf{E} = \epsilon \frac{\omega^2}{c^2} \mathbf{E} + \frac{4\pi i \omega}{c^2} \mathbf{j} \tag{8}$$

We consider a situation (Fig S1) where the monolayer medium spreads on the $xy$ plane with a current density $j_x = \tilde{\sigma}(\omega)\delta(z)E_x(x,t)$ that is driven by a propagating electric field on the $xz$ plane of the form $\mathbf{E} = (E_{0x}, 0, E_{0z})e^{i(\mathbf{k}\cdot\mathbf{r}-\omega t)}$. By evaluating the partial derivatives of **E**, the two components of the wave equation can be expressed as

$$ik_x \frac{\partial E_z}{\partial z} - \frac{\partial^2 E_x}{\partial z^2} - \epsilon \frac{\omega^2}{c^2} E_x = \frac{4\pi i \omega}{c^2} j_x \tag{9}$$

$$ik_x \frac{\partial E_x}{\partial z} + \left(k_x^2 - \epsilon \frac{\omega^2}{c^2}\right) E_z = 0 \tag{10}$$

where we have used $\partial/\partial x \to ik_x$ because the law of refraction requires that $k_x$ is conserved, while $k_z$ is not. Boundary conditions for the tangential and normal components of the field (red arrows) yield

$$E_x = (E_i - E_r)\cos\theta_i = E_t \cos\theta_t \tag{11}$$

$$\epsilon_s E_z|_{z=0^+} - E_z|_{z=0^-} = 4\pi \int_{0^-}^{0^+} \rho(z) dz \tag{12}$$

where we have used $\epsilon = \epsilon_s$ for the substrate and $\epsilon = 1$ for the vacuum. The charge density $\rho$ and the current density $j_x$ must satisfy the continuity equation $\partial\rho/\partial t + \nabla \cdot \mathbf{j} = 0$. Since they are driven by the same external field $E_x(x,t)$, we can then obtain a relation

$$\rho = j_x k_x / \omega \tag{13}$$

Equation (12) can now be evaluated by substituting $E_z$ from equation (10) and $\rho$ from equation (13), which yields

$$\frac{\epsilon_s}{k_s^2} \frac{\partial E_x}{\partial z^+} - \frac{1}{k_z^2} \frac{\partial E_x}{\partial z^-} = \frac{4\pi\tilde{\sigma}}{i\omega} E_x|_{z=0} \tag{14}$$

where the relations between $k_x, k_z$ and $k_s$ are shown in Fig S1. Note that the fields at the boundary are $E_x|_{z^+} = E_t e^{i(k_x x + k_s z)} \cos\theta_t$ and $E_x|_{z^-} = (E_i e^{i\mathbf{k}\cdot\mathbf{r}} - E_r e^{-i\mathbf{k}\cdot\mathbf{r}}) \cos\theta_i$. Substituting these will yield

$$\left(\frac{\epsilon_s}{k_s} + \frac{4\pi\tilde{\sigma}}{\omega}\right) E_t \cos\theta_t = \frac{1}{k_z} (E_i + E_r) \cos\theta_i \tag{15}$$

$$(E_i - E_r) \cos\theta_i = E_t \cos\theta_t \tag{16}$$

These are the two equations that will be used to obtain the reflectance, transmittivity, and absorptance of the monolayer. For convenience, we have moved equation (11) into (16).

At normal incidence, $k_s = \sqrt{\epsilon_s}\omega/c$ and $k_z = \omega/c$, hence the coefficients of amplitude reflection and transmission [8] can be simplified into

$$-r = -\frac{E_r}{E_i} = \frac{1 - n_s - \frac{4\pi\tilde{\sigma}}{c}}{1 + n_s + \frac{4\pi\tilde{\sigma}}{c}} \tag{17}$$

$$t = \frac{E_t}{E_i} = \frac{2}{1 + n_s + \frac{4\pi\tilde{\sigma}}{c}} \tag{18}$$

where $\tilde{\sigma} = \sigma_1 + i\sigma_2$ is the complex conductivity of the monolayer, and we have used $\sqrt{\epsilon_s} = n_s$ for an insulating substrate. Finally, we can obtain the reflectance $R$ and the transmittance $T$, as well as the absorptance $\alpha$ through the energy conservation $|r|^2 + n_s|t|^2 + \alpha = 1$ [8],

$$R = |r|^2 = \frac{\left(1 - n_s - \frac{4\pi\sigma_1}{c}\right)^2 + \left(\frac{4\pi\sigma_2}{c}\right)^2}{\left(1 + n_s + \frac{4\pi\sigma_1}{c}\right)^2 + \left(\frac{4\pi\sigma_2}{c}\right)^2} \tag{19}$$

$$T = n_s|t|^2 = \frac{4n_s}{\left(1 + n_s + \frac{4\pi\sigma_1}{c}\right)^2 + \left(\frac{4\pi\sigma_2}{c}\right)^2} \tag{20}$$

$$\alpha = \frac{4\left(\frac{4\pi\sigma_1}{c}\right)}{\left(1 + n_s + \frac{4\pi\sigma_1}{c}\right)^2 + \left(\frac{4\pi\sigma_2}{c}\right)^2} \tag{21}$$

The obtained $\alpha(\omega)$ is the absorptance of a monolayer medium deposited on an insulating substrate. The whole derivation already accounts for the out-of-phase back-reflected electric field that reduces the light intensity impinging on the monolayer. The above solutions can be expressed in terms of $\tilde{\epsilon}$ instead of $\tilde{\sigma}$ using the following relation

$$\tilde{\epsilon} = 1 + \frac{4\pi i \tilde{\sigma}/d}{\omega} \tag{22}$$

Note that $\tilde{\sigma}$ has different units in 2D (here) and in 3D; hence we keep the dielectric function dimensionless by introducing the monolayer thickness $d$. In order to convert these gaussian-unit equations into the SI-unit, we can use $4\pi \rightarrow 1/\epsilon_0$ where $\epsilon_0 (= 8.85 \times 10^{-12}$ F/m) is the vacuum permittivity.

## 4. Fitting analysis

To obtain the peak parameters shown in Figure 1 (fluence dependence) and Figure 4 (time dependence), we use a fitting expression that includes a Lorentzian function (exciton peak) and a second-order polynomial (background):

$$F(\omega) = I_0 \left( \frac{g_0}{\pi((\omega - \omega_0)^2 + g_0^2)} \right) + (A + B(\omega - \omega_1) + C(\omega - \omega_1)^2) \tag{23}$$

Here, $I_0, \omega_0$ and $g_0$ are the absorption peak intensity, energy and linewidth respectively, while $A, B, C$ and $\omega_1$ are the background constants and energy reference respectively. We first use this expression to fit the *equilibrium* absorption spectrum and record the fitting parameters. The obtained background parameters are fixed for subsequent fitting procedures in fluence-dependent and time-dependent spectra, and only the Lorentzian function parameters ($I_0, \omega_0$ and $g_0$) are allowed to vary.

## 5. Microscopic many-body computation

To calculate linear optical properties of monolayer WS$_2$ on a substrate under the influence of excited carriers, we combine first-principle $G_0W_0$ calculations with the solution of the semiconductor Bloch equations in screened-exchange-Coulomb-hole (SXCH) approximation for the two highest valence bands and the two lowest conduction bands as described in detail and applied to freestanding monolayer MoS$_2$ in Ref. [9]. In the following, we describe in detail how the previously used theory has to be augmented to properly take the substrate into account. We assume that the substrate mainly affects the internal Coulomb interaction and neglect its influence on the band structure, as we are only interested in relative shifts of the exciton resonance energy. Therefore, we derive the bare $U_{\alpha\beta}(q)$ and screened $V_{\alpha\beta}(q)$ Coulomb interaction *matrices* in the Wannier-orbital basis (with $\alpha, \beta \in [d_{z^2}, d_{xy}, d_{x^2-y^2}]$ ) for a freestanding WS$_2$ slab using the FLEUR and SPEX codes [10, 11]. As discussed in [12], macroscopic screening effects (like those arising from substrates) are described by the *leading* or *macroscopic* eigenvalue of the dielectric matrix. To access this quantity, we transform the full matrices $U_{\alpha\beta}(q)$ and $V_{\alpha\beta}(q)$ to their diagonal representations $U_d(q) = TU(q)T^*$ and $V_d(q) = TV(q)T^*$ using the eigenbasis $T$ of the bare interaction and define the diagonal dielectric function via $\varepsilon_d(q) = U_d(q)/V_d(q)$. Now each diagonal matrix is defined by its three eigenvalues. We fit the leading eigenvalues $U_1(q)$ and $\varepsilon_1(q)$ via

$$U_1(q) = \frac{e^2}{2\varepsilon_0 A} \frac{1}{q(1 + \gamma q + \delta q^2)} \tag{24}$$

$$\varepsilon_1(q) = \varepsilon_\infty(q) \frac{1 - \beta_1 \beta_2 e^{-2qh}}{1 + (\beta_1 + \beta_2) e^{-qh} + \beta_1 \beta_2 e^{-2qh}} \tag{25}$$

while all other elements ($U_2, U_3, \varepsilon_2$ and $\varepsilon_3$) are approximated by constant values given in Table S1.

| | U | | $\varepsilon$ | | T | | |
|---|---|---|---|---|---|---|---|
| | | | | | $d_{z^2}$ | $d_{xy}$ | $d_{x^2-y^2}$ |
| $U_2$ | 0.712 eV | $\varepsilon_2$ | 2.979 | | | | |
| $U_3$ | 0.354 eV | $\varepsilon_3$ | 2.494 | $T_1$ | +0.577 | +0.577 | +0.577 |
| $\gamma$ | 2.130 Å | $a$ | 3.989 Å$^{-2}$ | $T_2$ | +0.816 | -0.408 | -0.408 |
| $\delta$ | 0.720 Å$^2$ | $b$ | 30.19 | $T_3$ | 0 | -0.707 | +0.707 |
| $A$ | 2.939 Å$^2$ | $c$ | 5.447 Å | | | | |
| | | $h$ | 1.564 Å | | | | |
| | | $e$ | 4.506 | | | | |

**Table S1.** Fitting parameters to describe the diagonal bare interaction $U$, the corresponding eigenbasis $T$ and the diagonal dielectric function $\varepsilon$.

In Equation (24), $e$ is the elementary charge, $\varepsilon_0$ the vacuum permittivity, $A$ the unit cell area per orbital and $\gamma$ and $\delta$ are used to obtain optimal fits to the vacuum extrapolated ab initio data. In Equation (25) we introduced the parameters $\beta_i$ which are given by

$$\beta_i = \frac{\varepsilon_\infty(q) - \varepsilon_{sub.,i}}{\varepsilon_\infty(q) + \varepsilon_{sub.,i}} \tag{26}$$

Here, the dielectric constants of the substrate ($i = 1$) and superstrate ($i = 2$) are introduced. In order to describe the original ab initio data as close as possible we fit $\varepsilon_\infty(q)$ using

$$\varepsilon_\infty(q) = \frac{a + q^2}{\frac{a \sin qc}{qbc} + q^2} + e \tag{27}$$

and set $\varepsilon_{sub.,1} = \varepsilon_{sub.,2} = 1$. As soon as all fitting parameters are obtained (see Table S1) the screening of a dielectric environment can be included by choosing $\varepsilon_{sub.,1}$ or $\varepsilon_{sub.,2}$ correspondingly. In this paper, we use $\varepsilon_{sub.,1} = 10$ ($\varepsilon_{sub.,2} = 1$) which models the screening of the sapphire substrate. In Figure S2, we present the original ab initio data in combination with the resulting fits in the diagonal basis. In Figure S2c, we additionally show how the dielectric environment modulates the leading eigenvalue of the screening matrix. Using the latter, we can readily derive the fully screened Coulomb interaction $V_d(q) = U_d(q)/\varepsilon_d(q)$ including the screening effects of the dielectric environment. Finally, we make use of the transformation matrix $T$ to obtain the screened Coulomb interaction matrix in the orbital basis $V(q) = T^* V_d(q) T$.

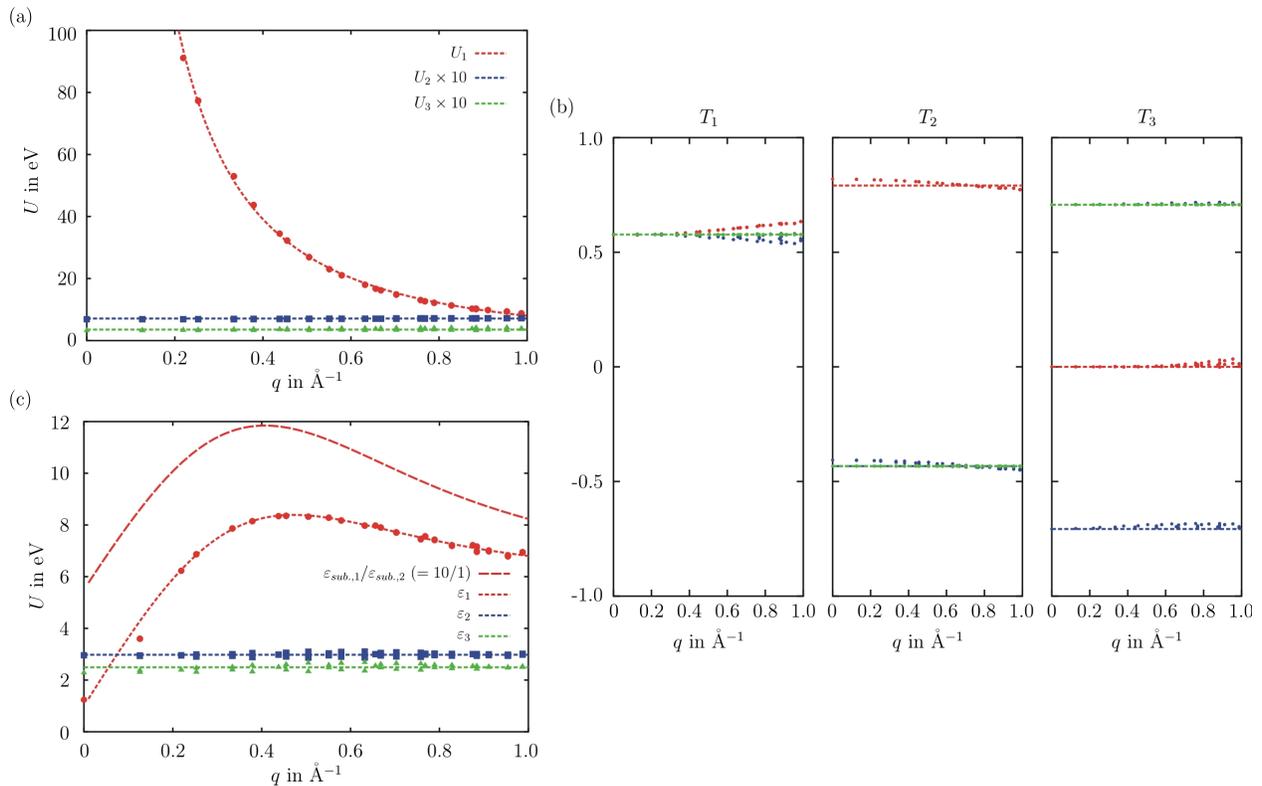

**Figure S2. (a)** Bare Coulomb matrix elements in its eigenbasis. Red dots, blue squares and green triangles correspond to the leading, second and third eigenvalue of $U(q)$ as obtained from ab initio calculations. Dashed lines show the corresponding fits using Equation (24) and Table S1. **(b)** Eigenvectors of the bare Coulomb matrix (from left to right corresponding to the leading, second and third eigenvalue). The corresponding vector elements of the $d_{z^2}$ (red), $d_{xy}$ (green) and $d_{x^2-y^2}$ (blue)

orbitals are shown. Dashed lines indicate constant fits as given in Table 1. **(c)** Matrix elements of the diagonal dielectric function. Markers indicate ab initio results and dashed lines show the fits using Equation (25) and Table S1. Next to the freestanding results we plot the leading eigenvalue of the dielectric matrix under the influence of a dielectric substrate with $\varepsilon_{sub,1} = 10$ (long dashes).

Besides the analytical description of the screened Coulomb matrix elements we make use of a Wannier-based tight-binding model to describe the electronic band structure (as obtained from $G_0 W_0$ calculations) of the WS$_2$ slab. To this end, we utilize the same Wannier-orbital basis as described before and derive a minimal three-band model describing the highest valence and two lowest conduction bands using the Wannier90 package [13]. Thereby we solely disentangle our target bands from the rest without performing a maximal localization in order to preserve the original W $d$-orbital characters. The latter is crucial for the subsequent addition of first and second order Rashba spin-orbit coupling following Ref. [14], which takes into account the large spin-orbit splitting in the conduction and the valence band K valleys. By using this computation approach, we can obtain the density dependence of gap shift, exciton binding energy, exciton peak shift and Bohr radius along the lines of Ref. [9] (Fig S3).

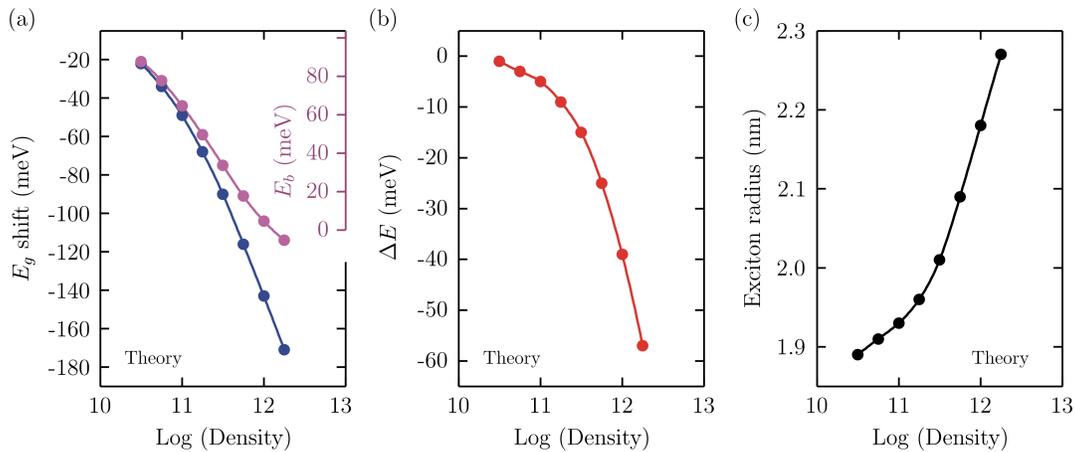

**Figure S3.** Computational results at increasing excitation density on the (a) gap narrowing $\Delta E_g$ and exciton binding energy $E_b$, (b) exciton shift $\Delta E$, and (c) exciton Bohr radius.

## 6. Exciton-exciton annihilation effect

At high carrier densities but below the Mott transition, plasma effects are small and the carriers predominantly form excitons. In this regime the effect of exciton-exciton annihilation has been discussed in the literature [15, 16]. With respect of a time delay of about 2 ps between excitation and measurement, we correct the estimation of the actual exciton density by the reduction due to this process. In order to estimate the dissipation rate, we studied the exciton bleaching decay upon photoexcitation with 3.16 eV pump pulse. Fig S4 shows two time-traces of $-\Delta\alpha$ at the A exciton absorption peak with pump fluences of 14.6 μJ/cm$^2$ and 4.4 μJ/cm$^2$, where the measured data is shown by the open circles. These pump fluences correspond to excitation densities of $n_0 = 4.3 \times 10^{12}$ cm$^{-2}$ and $1.3 \times 10^{12}$ cm$^{-2}$.

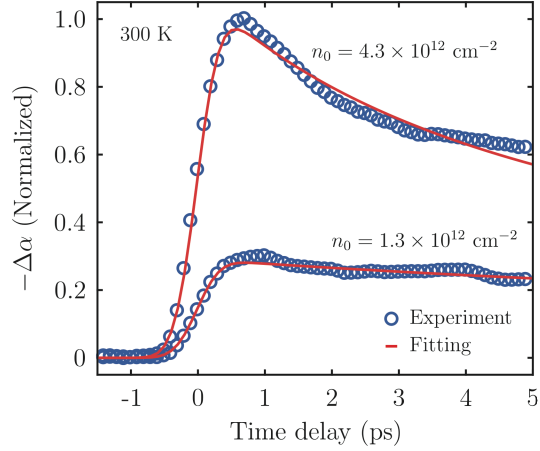

**Figure S4.** Time-traces at different pump fluences to study the exciton annihilation effect on the density. The upper and lower curves correspond to pump fluences of 14.6 µJ/cm² and 4.4 µJ/cm² (open circles are measured data points, red curves are fitting lines). These pump fluences correspond to excitation densities of $n_0 = 4.3 \times 10^{12}$ cm$^{-2}$ and $1.3 \times 10^{12}$ cm$^{-2}$.

Exciton-exciton annihilation can be described by differential equation

$$\frac{dn}{dt} = -kn^2 \tag{28}$$

where $n$ is the exciton density at time $t$, and $k$ is the annihilation rate. This differential equation has a solution

$$n(t) = \frac{n_0}{1 + kn_0 t} \tag{29}$$

where $n_0$ is the initial excitation density. Through global fitting of the two time-traces we obtain an annihilation rate $k = 0.04 \pm 0.01$ cm²/s, where the red curves show the fitting lines. The obtained value is consistent with those reported for CVD-grown monolayer WS$_2$ of 0.08-0.10 cm²/s [17, 18], within an order of magnitude. The uncertainty in the fitting curve at high fluence is a common observation due to the 3.16 eV above-gap excitation and attributed to the fast cooling process [17]. Note that the annihilation rate is dependent on particular substrate used and on the as-grown sample quality. Exfoliated monolayer WS$_2$ shows greater annihilation rate [16], which is also discussed in Reference [17]. By using the obtained annihilation rate, we renormalize the density on Fig 3a. At the max fluence we used, the actual density reaches 70% at 2 ps, while at low fluence the actual density reaches 90% at 2 ps.

## 7. Heat capacity and estimated temperature

In the later parts of our experiments that involve variations in time delays and pump fluence, we used pump photon energy of 3.16 eV, which is much higher than the lowest excitation energy (A exciton) of 2.00 eV. This means that, for every e-h pair excited in the pumping process, there will be about 1.16 eV of excess energy after relaxation into the A exciton. In monolayer $WS_2$, as is also the case for most materials, the relaxation processes are dominated by electron-electron (e-e) and electron-phonon (e-ph) scatterings. Typically, the timescale of e-e thermalization is about 10−100 fs, while the e-ph thermalization is about 1 ps. This means, the excess energy will be first distributed among the electrons to form a hot exciton gas, followed by heat transfer into the lattice. Here, we want to estimate the electronic temperature $T_e$ and the lattice temperature $T_l$ by calculating the corresponding heat capacities. Note that these temperatures will be the upper limits of what we expect from the system because the heat transfer to the substrate is known to be very effective in 2D systems, with a timescale of about 2 ps [19].

The electronic heat capacity $C_e$ (per area) can be expressed as

$$C_e(T) = \int \frac{\partial f(\varepsilon, \mu)}{\partial T} \varepsilon D(\varepsilon) d\varepsilon \tag{30}$$

where $f$ is the occupation number of states, $\mu$ is the chemical potential, and $D(\varepsilon)$ is the density of states (per area) in the range of energy $\varepsilon$ and $\varepsilon + d\varepsilon$. The low-energy excitations in monolayer $WS_2$ constitute of excitons with various spin combinations in the two valleys (Fig S5).

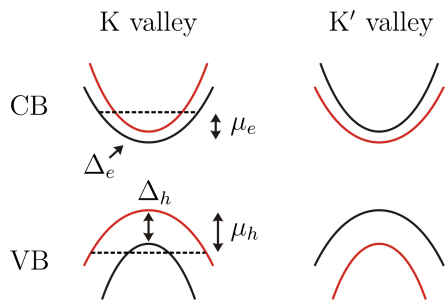

**Figure S5.** Schematic electronic band structure (one-particle picture) of monolayer $WS_2$ at the K and K' valleys. The CB consists of two electron bands at each valley separated by a spin-splitting gap of $\Delta_e \sim 30$ meV, and the VB consists of two hole bands at each valley separated by a spin-splitting gap of $\Delta_h \sim 400$ meV [20, 21]. The chemical potentials ($\mu_e$ and $\mu_h$) of the photoexcited sample are measured from the tip of the relevant bands.

Here, we will consider the contributions of the electrons and holes to the electronic heat capacity separately. In this way, we can account for the spin-valley degeneracy by assuming parabolic energy dispersion for each band as $\varepsilon_n = \hbar^2 k^2 / 2m_n + \Delta_n$, where $m_n$ and $\Delta_n$ are the effective mass and the gap of band $n$ (with specific spin-valley index). The density of states can then be expressed as $D(\varepsilon) = \sum_n \theta(\varepsilon - \Delta_n) m_n / 2\pi\hbar^2$, while the occupation number is $f(\varepsilon, \mu) = [\exp((\varepsilon - \mu)/kT) + 1]^{-1}$. The chemical potential, which depends on excitation density $n$ and temperature, has an important role to keep the number of electrons and holes equal ($n_e = n_h = n$). In this quasi-equilibrium condition, where we have intentionally photo-injected the carriers into the system, the chemical potentials $\mu_e$ and $\mu_h$ are measured from the bottom of the conduction band (CB) and the top of the valence band (VB), respectively, as are also the case for the kinetic energies $\varepsilon_n$. Variations of these chemical potentials with temperature can be followed from the conservation of particle's number density $n = \int f(\varepsilon, \mu) D(\varepsilon) d\varepsilon$. In the case of electrons, this gives

$$n_e/n_T = \ln\left[\left(1 + e^{\mu_e/kT}\right)^2 \left(1 + e^{(\mu_e - \Delta_e)/kT}\right)^2\right] \tag{31}$$

and the holes,

$$n_h/n_T = \ln\left[\left(1 + e^{\mu_h/kT}\right)^2 \left(1 + e^{(\mu_h - \Delta_h)/kT}\right)^2\right] \tag{32}$$

where $n_T = \langle m_n \rangle kT / 2\pi\hbar^2$ is called the thermal quantum density. Hence, for a given excitation density, we can compute the chemical potentials as a function of temperature. Finally, by using the above equation, we can calculate the electronic heat capacity as a function of temperature (Fig S6a). The actual electronic heat capacity is expected to be larger than this because of the higher-lying bands which we have ignored in the present calculations.

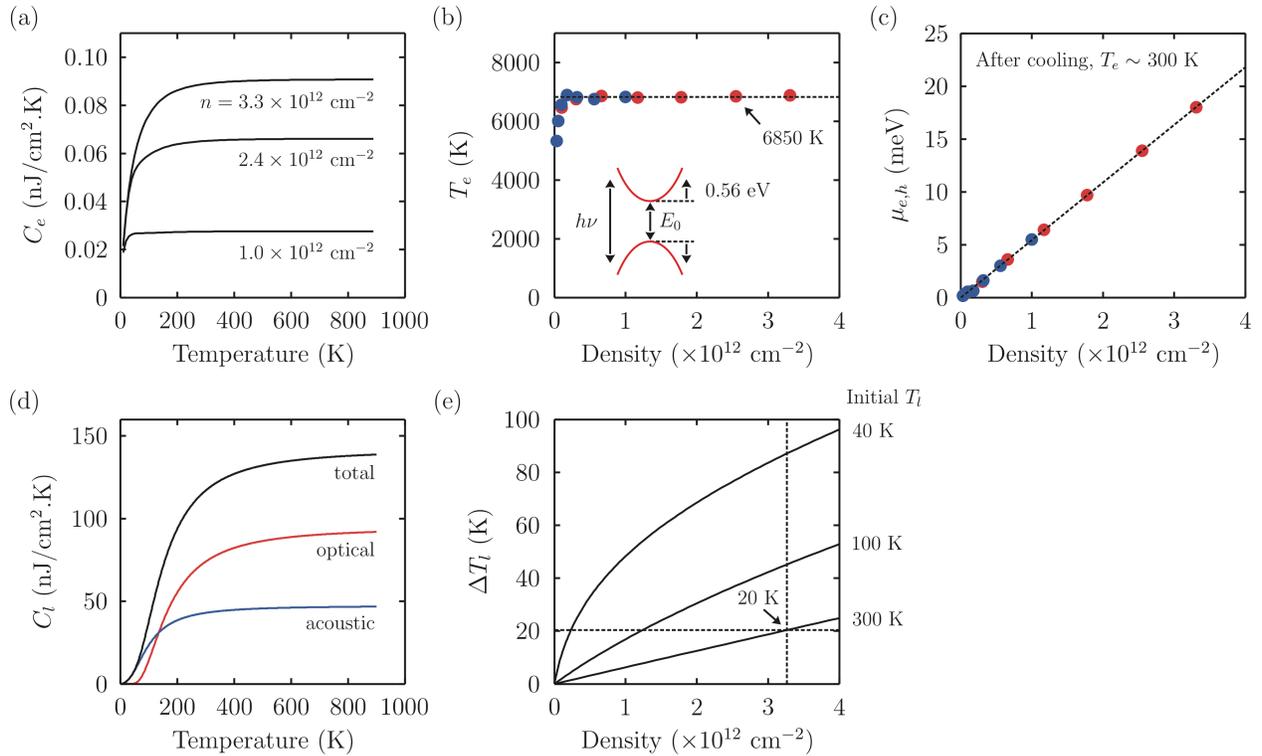

**Figure S6.** Estimating the electronic and lattice temperatures after photoexcitation by calculating the heat capacities and the absorbed excess energy. **(a)** Electronic heat capacity $C_e(T)$ of monolayer WS$_2$ by considering the lowest e-h excitations at K,K' valleys. **(b)** $T_e$ vs $n$. **(c)** $\mu_{e,h}$ vs $n$. **(d)** Lattice heat capacity $C_l(T)$ by considering three acoustic and six optical phonon modes. **(e)** $\Delta T_l$ vs $n$. Here, the electrons and lattice are treated separately.

As shown in Fig S6a, $C_e$ increases rapidly at low temperatures and saturates at higher temperatures. Note that heat capacity is proportional to the number of electrons that can store the thermal energy. At low temperatures (quantum regime), only the electrons around the Fermi level can contribute to $C_e$. At higher temperatures (classical regime), electrons are more sparsely distributed across different energies, hence more electrons can contribute to $C_e$ until it reaches a saturation value where all of the electrons are involved. The transition between quantum and classical regimes should happen at temperature reaching the electron-hole chemical potential, $k_bT \sim \mu_{e,h}$. This is also consistent with the fact that in classical regime $C_e$ is proportional to the excitation density $n$.

We can now estimate the rise of temperature upon photoexcitation in monolayer WS$_2$ by using

$$\Delta Q = n\Delta E = \int C(T)dT \tag{33}$$

where $\Delta Q$ is the absorbed energy density, $n$ is the photoexcited pair density and $\Delta E$ is the excess energy per pair (1.13 eV). Note that in monolayer WS$_2$, the usual (singlet) A exciton is slightly higher by $\Delta_e = 30$ meV as compared to the (triplet) A exciton. In the first few hundreds of femtoseconds after photoexcitation, most of the excess energy is redistributed among the electrons. By using the obtained $C_e(T)$ and $\Delta Q$, we can calculate $T_e$ for a given excitation density as shown in Fig S6b. The results show that, except at very low densities, the electronic system reaches a constant $T_e = 6850$ K at all densities. In fact, this result can be understood if we assume that every photoexcited charge carrier (electron or hole) carries an excess photon energy of $(h\nu - E_0)/2 = 0.56$ eV, which will be stored as their thermal energy $k_bT_e$. This will amount to $T_e \sim 6500$ K regardless of the density, and this is consistent with the above results. So, if we were to use higher photon energy, $T_e$ would increase correspondingly. Therefore, in ideal condition, the fluence dependence data in Fig 1 (main text) should correspond to the same temperature of 6850 K at all fluences. However, in reality the transient electronic temperature is usually much smaller (typically $T_e > 1000$ K) due to rapid thermalization with phonons that we discuss below.

In short timescale, the thermal energy will be distributed to the lattice (~1 ps) or substrate (~10-1000 ps), and $T_e$ will decrease back to 300 K. In this situation, we can estimate the chemical potential $\mu_{e,h}$ of the charge carriers for a given excitation density. Figure S6c shows that $\mu_{e,h}$ increases linearly with density, and the chemical potentials are the same for electrons and holes because they have similar effective mass in monolayer WS$_2$ ($m_e = m_h = 0.44\ m_0$, [22]). Such a linear increase in $\mu(n)$ is quite expected due to the constant density of states $D(\varepsilon)$ in an ideal 2D system. Note that, at the excitation density that we used in the experiment, $\mu_{e,h}$ is still lower than the spin-splitting of the two bands ($\Delta_e = 30$ meV, $\Delta_h = 400$ meV). So, for much higher excitation densities, we would expect the $\mu_{e,h}$ vs $n$ slope to be lowered into half as it approaches $\mu_{e,h} \sim 30$ meV.

For the lattice heat capacity $C_l$, we calculated separately the contribution from the acoustic and optical phonon modes. Monolayer WS$_2$ has three atoms in the unit cell, with three acoustic and six optical modes. By taking the average optical phonon energy as $\hbar\omega_0 = 45$ meV [23], we can estimate its contribution to $C_l$ per unit cell as

$$C_{op}(T) = \frac{\partial}{\partial T}\left(\frac{6\hbar\omega_0}{e^{\hbar\omega_0/kT} - 1}\right) \tag{34}$$

For the acoustic phonon contributions (per unit cell), we can use the 2D Debye model which gives

$$C_{ac}(T) = 6k_b\left(\frac{T}{\Theta}\right)^2 \int_0^{x_D} \frac{x^3 e^x}{(e^x - 1)^2} \tag{35}$$

where the Debye temperature is defined as $\Theta = (\hbar v/k_b)\sqrt{4\pi/A_{cell}}$ and the Debye cutoff is $x_D(T) = \Theta/T$. We calculated the Debye temperature ($\Theta = 460$ K) by using an average sound velocity $v$ of about $5\times10^3$ m/s [23], and an area per unit cell $A_{cell}$ of $8.46\times10^{-16}$ cm$^2$ with a lattice constant of $a = 3.13$ Å [23]. Finally, the total lattice heat capacity can be calculated, which is shown in Fig S6d.

Figure S6d shows that $C_l(T)$ increases rapidly at low temperatures and saturates to its Dulong-Petit value at higher temperatures. Note that $C_l$ is about 1000 times larger than $C_e$, and this means for a given $n$ the temperature increase $\Delta T_l$ would be much smaller than $\Delta T_e$. Figure S6e shows $\Delta T_l$ as a function of $n$ for different initial lattice temperatures $T_l$. Since in our time-dependent experiment we used $n = 3.3\times10^{12}$ cm$^{-2}$ at $T_l = 300$ K (Fig 4, main text), $\Delta T_l$ increases by only about 20 K. Unlike the electronic system where the excess photon energy is redistributed evenly to the charge carriers that results in a constant $C_e(n)$, the lattice will absorb the total excess energy from all of the charge carriers. Hence, $C_l$ increases with $n$.

Now, if we allow a strong e-ph coupling to direct the absorbed energy into the lattice, we can merely use the total lattice heat capacity to estimate the temperature of the system. This is because $C_l \gg C_e$. As we can see from Fig S6b and S6e, with excitation density of $3.3\times10^{12}$ cm$^{-2}$ (Fig 4, main text), the electronic temperature could reach $T_e > 1000$ K for a short while until it cools down to share a common temperature with the lattice to about 320 K. In fact, we must also consider the heat transfer to the thick substrate, which effectively plays a role as the thermal reservoir at 300 K. As a result, the actual temperatures should be much lower than what we have estimated in this analysis, similar to what have been observed in suspended vs supported graphene [19].